\documentstyle[12pt]{article}

\def\la{\mathrel{\mathchoice {\vcenter{\offinterlineskip\halign{\hfil
$\displaystyle##$\hfil\cr<\cr\sim\cr}}}
{\vcenter{\offinterlineskip\halign{\hfil$\textstyle##$\hfil\cr<\cr\sim\cr}}}
{\vcenter{\offinterlineskip\halign{\hfil$\scriptstyle##$\hfil\cr<\cr\sim\cr}}}
{\vcenter{\offinterlineskip\halign{\hfil$\scriptscriptstyle##$\hfil\cr<\cr
\sim\cr}}}}}

\input epsf
\fussy

\begin{document}

\begin{center}
\Large
{\bf Azimuthal Correlations in the Target Fragmentation Region\\
of High Energy Nuclear Collisions\\}
\vspace{1cm}

\normalsize
T.C.\,Awes$^{e}$,
D.\,Bock$^{d}$,
R.\,Bock$^{a}$,
G.\,Clewing$^{d}$,
S.\,Garpman$^{c}$,
R.\,Glasow$^{d}$,
H.\AA.\,Gustafsson$^{c}$,
H.H.\,Gutbrod$^{a}$,
G.\,H\"{o}lker$^{d}$,
P.\,Jacobs$^{b}$,
K.H.\,Kampert$^{d,1}$,
B.W.\,Kolb$^{a}$,
Th.~Lister$^{d}$,
H.\,L\"{o}hner$^{h}$,
I.\,Lund$^{h}$,
F.E.\,Obenshain$^{e}$,
A.\,Oskarsson$^{c}$,
I.\,Otterlund$^{c}$,
T.\,Peitzmann$^{d}$,
F.\,Plasil$^{e}$,
A.M.\,Poskanzer$^{b}$,
M.\,Purschke$^{a}$,
H.G.\,Ritter$^{b}$,
R.\,Santo$^{d}$,
H.R.\,Schmidt$^{a}$,
T.~Siemiarczuk$^{a,2}$,
S.P.\,S{\o}rensen$^{e,i}$,
K.\,Steffens$^{d}$,
P.\,Steinhaeuser$^{a,1}$,
E.\,Stenlund$^{c}$,
D.\,St\"{u}ken$^{d}$,
and G.R.\,Young$^{e}$\\
\vspace{0.5cm}
{\bf WA80 Collaboration}\\
\vspace{0.5cm}
\small
\em
\begin{tabular}{ll}
a.  & Gesellschaft f\"{u}r Schwerionenforschung, D-64220 Darmstadt, Germany\\
b.  & Lawrence Berkeley Laboratory, Berkeley, California 94720, USA\\
c.  & University of Lund, S-22362 Lund, Sweden\\
d.  & University of M\"{u}nster, D-48149 M\"{u}nster, Germany\\
e.  & Oak Ridge National Laboratory, Oak Ridge, Tennessee 37831, USA\\
h.  & KVI, University of Groningen, NL-9747 AA Groningen, Netherlands\\ 
i.  & University of Tennessee, Knoxville, Tennessee 37996, USA
\vspace{0.5cm}
\end{tabular}
\end{center}

\begin{abstract}
\noindent Results on the target mass dependence of 
proton and pion pseudorapidity distributions and of their azimuthal 
correlations in the target rapidity range $-1.73 \le \eta \le 1.32$ 
are presented.  The data have been taken with the Plastic-Ball 
detector set-up for 4.9~GeV p\,+\,Au collisions at the Berkeley 
BEVALAC and for 200 $A\cdot$GeV/$c$ p-, O-, and S-induced reactions 
on different nuclei at the CERN-SPS.  The yield of protons at 
backward rapidities is found to be proportional to the target mass.  
Although protons show a typical ``back-to-back'' correlations, a 
``side-by-side'' correlation is observed for positive pions, which 
increases both with target mass and with impact parameter of a 
collision.  The data can consistently be described by assuming 
strong rescattering phenomena including pion absorption effects in 
the entire excited target nucleus.
\end{abstract}
\normalsize

\begin{twocolumn}
\newpage
\noindent
The investigation of pion and baryon spectra and collective flow 
phenomena at relativistic energies is a well established field of 
nuclear research [1-3].  It has been addressed theoretically more 
than 20 years ago [4,5] and was investigated for the first time at 
the BEVALAC more than 10 years ago [6].  Recently, the study of the 
{\em interaction\/} of these particle species among themselves 
within the nuclear medium and the possible formation of 
$\Delta$-matter has gained renewed attraction [7-12].  At 
relativistic energies, i.e.\ at bombarding energies around 1 
$A\cdot$GeV, the most important process involving nucleons {\em 
and\/} pions is the excitation of the $\Delta(1232)$ resonance.  As 
the fate of a pion produced at these beam energies is governed by 
the reactions $\pi NN \to \Delta N \to NN$ and $\pi N \to
\Delta \to \pi N$, one obviously should focus onto observables,
where pion absorption and rescattering plays a role.  Observables 
influenced by these processes \linebreak[4] should be the pion 
abundance itself as well as their energy and azimuthal 
distributions with respect to the reaction plane.  The collision 
geometry of heavy-ion reactions delivers a ``gauge'': once the 
reaction plane is known, one can find regions (in 3-dimensional 
coordinate space) where pions can escape the reaction zone with 
either minimal or maximal reinteraction in baryonic spectator 
matter [13].

In this letter we shall present results from proton-nucleus and 
nucleus-nucleus reactions both at relativistic (4.9 GeV) as well as 
at ultrarelativistic energies (200 $A\cdot$GeV/$c$).  It has been shown 
[14] that the target fragmentation region at ultrarelativistic 
energies features similar characteristics as the central rapidity 
zone at relativistic energies, i.e.\ by restricting the 
investigation to $y \la 0$ one can expect to cover the resonance 
regime as discussed above.  We shall investigate azimuthal 
correlations both for positive pions and protons and study their 
dependence on the geometry of the reaction system.

The data were taken employing the Plastic Ball detector [15] at the 
Berkeley BEVALAC and at the CERN-SPS in the WA80 experiment.  The 
Plastic Ball is a modular, azimuthally symmetric array of $\Delta 
E$-$E$ telescopes covering the polar range from $160^\circ$ to 
$30^\circ$ in the laboratory.  Full particle identification is 
achieved for particles stopped in the $E$ counter, thus limiting the 
energy range of accepted light baryons to 40 MeV $\le E_{\rm kin}/A 
\le 240$~MeV and of positive pions to 20 MeV $\le E_{\rm kin}^\pi 
\le 120$~MeV.  The pseudorapidity coverage of \linebreak[4] $-1.73 \le
\eta \le 1.32$ is ideally suited to study the target fragmentation
region at ultrarelativistic energies.  If not mentioned 
differently, the data were analyzed under minimum bias trigger 
conditions.

\begin{figure}[t]
\centerline{\epsfxsize=7.0cm \epsfbox{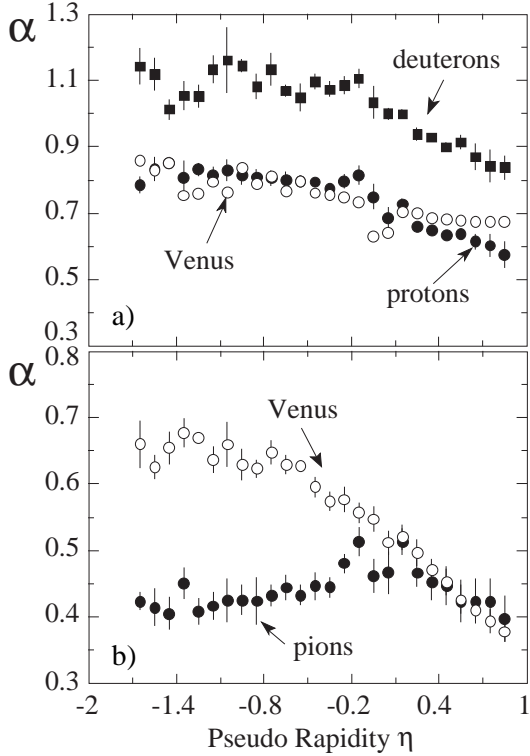}}
\caption[x]{\small
(a): Target mass dependence of protons (filled circles)
and deuterons (filled squares) as a function of $\eta$ for 
p\,+\,$A$ reactions at 200\,GeV/$c$ incident momentum.  The target 
dependence is parameterized as $dN/d\eta~\propto~A^{\alpha(\eta)}$.  
The open circles show the result of VENUS simulations filtered to 
the experimental acceptance.\\
(b): Same as in (a) but for positive pions (filled circles).
The open circles show again the result of VENUS simulations.}
\end{figure}

An investigation of the target mass dependence of the proton, 
deuteron, and pion yields is depicted in Fig.\ 1 for 200\,GeV/$c$ 
p\,+\,$A$ reactions ($A \equiv $ C, Al, Cu, Ag, Au).  The 
individual yields have been parameterized as $dN/d\eta(A)\propto 
A^{\alpha(\eta)}$ and the exponent $\alpha$ is plotted as a 
function of $\eta$ .  The backward protons (Fig.\ 1a) show a clear 
trend towards a value of $\alpha = 1$, while backward deuterons 
even exceed $\alpha = 1$.  A previous analysis accepting {\em 
all\/} charged particles has provided very similar results [16].

The stronger target mass dependence of deuteron production is 
qualitatively under- \linebreak[4] stood within the coalescence 
picture, where $\rho_{d} \propto \rho_{p} \cdot \rho_{n}$ [17].
Based on the experimental observation of
$\rho_{p} \propto A$, and assuming $\rho_{p} \approx 
\rho_{n}$, a dependence of deuteron production $\rho_{d} \propto
A^{2}$ is inferred, i.e.\ the {\em average} $A$-dependence of 
deuteron production scales with $A^{\alpha}$ with $\alpha > 1$.

In any event, the observed $A$-dependence of baryons is a clear 
indication of an homogeneous excitation of the entire target 
nucleus.  The experimental findings for baryons are well reproduced 
by VENUS 4.12 simulations.  VENUS [18] is a string model which 
treats rescattering of secondary particles in a rather simplistic 
way: as two objects approach each other below a critical distance 
they fuse with subsequent isotropic decay.  This corresponds, for 
the case of $NN$ interactions, to elastic $NN$ scattering.

The situation is quite different for pions as shown in Fig.\ 1b.  
Here $\alpha$ stays at about 0.4 for all values of $\alpha$, while, 
in contrast, the simulated pion yields from VENUS exhibit an 
increasing value of $\alpha$ for decreasing values of $\eta$.  This 
behavior might be taken as an indication that the large resonance 
cross section (which is not taken into account in VENUS) plays an 
important role to decrease the pion yield via absorption as the 
target mass increases.

More information about possible absorption and rescattering effects 
can be obtained from azimuthal particle correlations.  To search 
for such kind of effects between protons or between pions in the 
target rapidity range, a correlation function $C(\Delta\varphi)$ is 
constructed as follows:

\begin{displaymath} 
C(\Delta\varphi) \equiv
\frac{dN} {d(\Delta\varphi)} \,\, ; \, {\rm where} 
\end{displaymath}

\begin{displaymath}
\Delta\varphi = \arccos \left(
\frac{\vec{Q}_{ \rm back} \cdot \vec{Q}_{ \rm forw}}
{\vert \vec{Q}_{ \rm back}\vert \cdot \vert \vec{Q}_{ \rm forw} \vert}  \right)
\;\; {\rm  with}
\end{displaymath}

\begin{displaymath}
\vec{Q}_{ \rm back} \equiv \sum_{y <    y_{0}} \vec{p}^{\,\,i}_{\perp} \;\; {\rm and} \;\; 
\vec{Q}_{ \rm forw} \equiv \sum_{y \geq y_{0}} \vec{p}^{\,\,i}_{\perp}
\end{displaymath}  

\noindent The value of $y_{0}$ is chosen as 0.2. This is guided
by the experimental fact that the target rapidity distribution of 
protons peaks at this value [19] and follows the idea of a target 
``fireball'' moving with a rapidity $y_{0}$.  The influence of the 
actually chosen value of $y_{0}$ to the experimental results, 
furthermore, has been checked by varying $y_{0}$ in a reasonable 
range $0.1 \le y_{0} \le 0.3$.  Within these limits no significant 
change of the correlation function has been observed.

Essentially, $C(\Delta\varphi)$ measures whether the particles in 
the backward and forward hemispheres of the target fireball are 
preferentially emitted ``back-to-back'' $(\Delta\varphi = 
180^{\circ})$ or ``side-by-side'' $(\Delta \varphi = 0^{\circ})$, 
meaning on the opposite or on the same side of the reaction plane, 
respectively.

Figure 2 shows the experimental correlation function 
$C(\Delta\varphi)$ under minimum trigger bias conditions for 
4.9\,GeV (top) and 200\,GeV/$c$ (bottom) protons impinging on a Au 
target.  The lefthand and righthand figures present 
$C(\Delta\varphi)$ for protons and positive pions, respectively.

Both, for protons and for pions a clear correlation is observed, 
but of opposite direction.  To quantify these experimental results, 
the data were fitted by $C(\Delta\varphi) \propto 1 + \xi
\cos(\Delta\varphi)$ and the strength of the correlation is defined 
as

\begin{displaymath}
\zeta \equiv \frac{C(0^{\circ})} {C(180^{\circ})}  =
\frac{1 + \xi} {1 - \xi} .
\end{displaymath}

\begin{figure}[tb]
\centerline{\epsfxsize=7.6cm \epsfbox{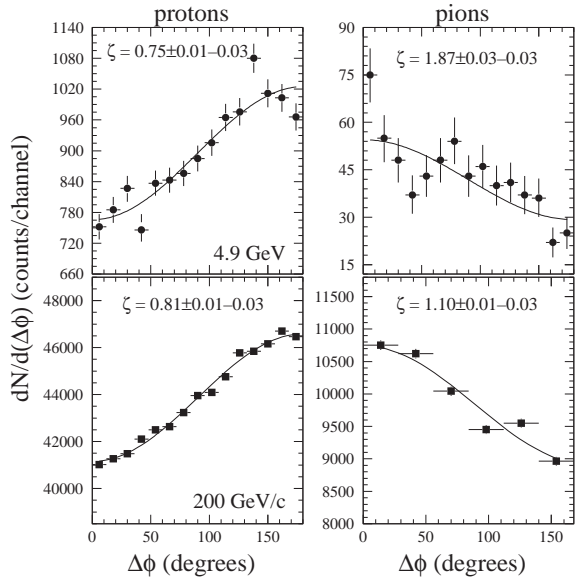}}
\caption[x]{\small The correlation function $C(\Delta\varphi)$ as
described in the text for p\,+\,Au collisions at 4.9 GeV (top) and
200 GeV/$c$ (bottom). The lefthand and righthand panels present
$C(\Delta\varphi)$ for protons and positive pions, respectively.}
\end{figure}

As can be seen, one observes $\zeta <1$ for protons and $\zeta > 1$ 
for pions, meaning that protons are preferentially emitted 
back-to-back, while pions are emitted side-by-side with respect to 
$y_{0}$.  Detector asymmetries, which \linebreak[4]
might cause an artificial 
side-by-side correlation, have been studied carefully by 
investigating the azi\-muth\-al distributions of $\vec{Q}_{\rm 
back}$ and $\vec{Q}_{\rm forw}$ individually.  The observed maximum 
deviations from azimuthal symmetry allow to set a limit to the 
influence of the correlation function by +0.03 at most.  This 
error is indicated in the values of $\zeta$.

The back-to-back emission of protons can be understood as resulting 
from (local) transverse momentum conservation.  This interpretation 
is supported also by results from string models like VENUS.  On the 
other hand, the side-by-side correlation of pions can naturally be 
explained based on the picture that pions, which are created in a $ 
b \neq 0$~fm collision either suffer rescattering or even complete 
absorption in the target spectator matter.  Both processes will 
result in a relative depletion of pions in the geometrical 
direction of the target spectator matter and hence will cause an 
azimuthal side-by-side correlation as observed in the experimental 
data.  While for a $b \approx 0$~fm collision the emission 
directions of pions with respect to the reaction plane are expected 
to become azimuthally symmetric, the azimuthal distributions are 
expected to become more and more asymmetric as the impact parameter 
increases.

This hypothesis is clearly supported by the observed target mass 
and the impact parameter dependence of the correlation.  The latter 
dependence has been determined both for p-induced and for 
heavy-ion induced reactions by applying cuts to the transverse 
energy measured event-by-event with the MId-RApidity Calo\-rimeter 
(MIRAC) [20].  In all systems, a plateau in $dE_{T}/d\eta$ as a 
function of $E_{T}$ is observed with a steep fall-off towards 
higher transverse energies.  To select two regions of impact 
parameters, the events were divided into two bins with $E_{T}$ 
smaller and larger than $2/3 \cdot E_{T}^{\rm max}$.  These events 
are denoted as ``peripheral'' and ``central'', respectively.  More 
stringent cuts on $E_{T}$, i.e.\ on the impact parameter or on the 
violence of the collision, were found to result in intolerable 
large statistical uncertainties of the extracted asymmetry 
parameter $\zeta$.  These uncertainties were on the one hand caused 
by the low multiplicity of pions in the target rapidity region (low 
$E_{T}$) and on the other hand by the small number of events 
passing the software cut at high $E_{T}$.

\begin{figure}[t]
\centerline{\epsfxsize=7.6cm \epsfbox{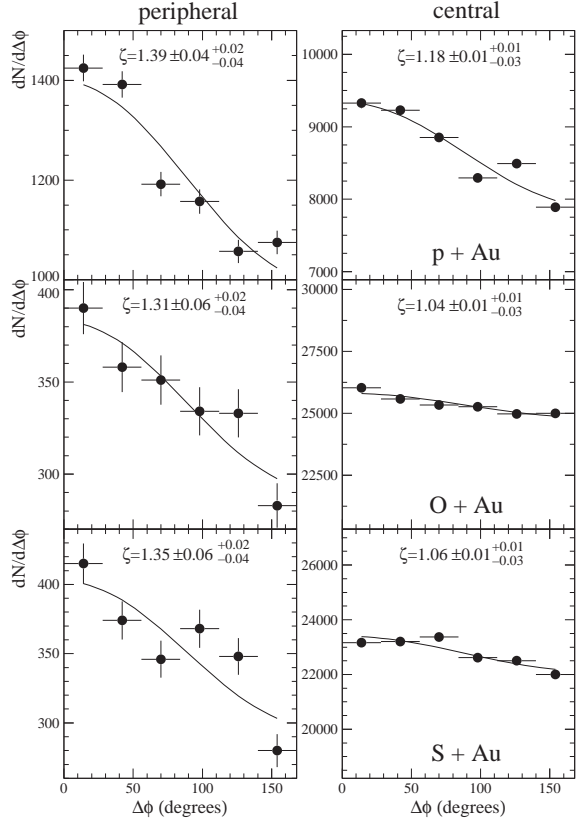}}
\caption[x] {\small The correlation function $C(\Delta\rho)$ for
positive pions from p\,+\,Au (top) and O\,+\,Au (middle), and 
S\,+\,Au (bottom) reactions at 200 GeV/$c$ taken under minimum bias 
conditions.  The lefthand and righthand panels are selections of 
peripheral and central events, respectively.}
\end{figure}

Figure 3 shows the correlation function for positive pions from 
p\,+\,Au, O\,+\,Au, and S\,+\,Au reactions both for peripheral and 
central collisions, respectively.  The correlation function or, 
equivalently, the correlation \linebreak[4] strength parameter 
clearly shows a behavior as one would expect from the rescattering 
and absorption picture in spectator matter: the strength of the 
correlation is enhanced for peripheral collisions and almost 
vanishes for central O\,+\,Au and S\,+\,Au collisions.  
Furthermore, the azimuthal asymmetry appears to be stronger for 
p-induced reactions than for O- and S-induced reactions.  Both 
variations can be interpreted by the different amounts of spectator 
matter available for secondary interactions.  The systematic error 
of the asymmetry parameter $\zeta$ is in peripheral collisions 
mainly caused by relative uncertainties in selecting the same 
`violence' of the collision for different reaction systems by 
applying software trigger cuts to the associated transverse energy 
$E_{T}$, and is in central collisions dominated by azimuthal 
asymmetries in the detector response itself (see above).  The 
residual azimuthal correlation observed in central O\,+\,Au and 
S\,+\,Au reactions can thus to a large extend be explained by 
detector asymmetries.

\begin{figure}[tb]
\centerline{\epsfxsize=7.6cm \epsfbox{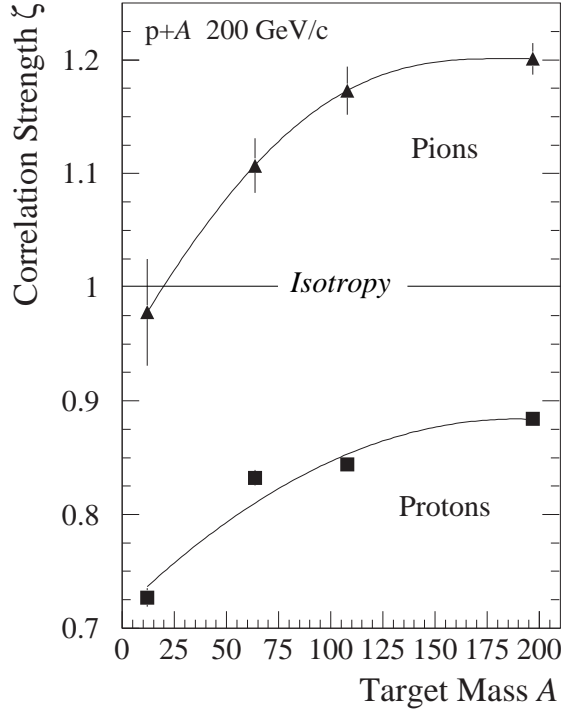}}
\caption[x] {\small Target mass dependence of the correlation
strength parameter $\zeta$ for both $\pi^{+}$ and protons from
minimum bias proton induced reactions at 200 GeV/$c$. Only 
statistical errors are shown. The curves are to guide the eye.}
\end{figure}

The dependence of the correlation coefficients on the target mass 
in minimum bias data is summarized in Fig.\ 4 for protons and 
positive pions.  The curves through the data points are to guide 
the eye.  Again, we stress the interpretation within the 
geometrical absorption picture: as the target mass increases the 
side-by-side asymmetry increases for the pions due to the 
increasing amount of matter in their path.  In contrast, the 
back-to-back asymmetry of protons tends to vanish.  This is also 
expected, because correlations due to momentum conservation, as 
discussed above, become weaker for higher event multiplicities, 
i.e.\ for heavier system.

The observations presented in this Letter are consistent with a 
recent investigation of charged pion flow [10] in symmetric 
heavy-ion collisions at SIS energies.  Here, anisotropic pion flow 
relative to the reaction plane was found (``pion squeeze-out'') and 
absorption of pions in the reaction plane was conjectured to be the 
cause for the anisotropy.  Due to the different reaction geometry 
in symmetric systems, the spectator matter located in the reaction 
plane causes a stronger absorption and rescattering of pions in 
plane than out of plane.

Another investigation of large angle two-particle correlations, 
carried out at the 3.6 $A\cdot$GeV C-beam in Dubna [21] showed a 
back-to-back pion correlation for a light target (Al), no 
correlation for a medium target (Cu), and a side-by-side 
correlation for a heavy target (Pb).  For protons and deuterons, a 
back-to-back correlation was observed for all targets.  Again, 
these results appear to be consistent with our observed proton 
and pion azimuthal correlations and with the variation of the pion 
correlation when going from a C to a Au target.

In summary, we have found conclusive experimental evidence for pion 
absorption and rescattering at target rapidities in 
ultrarelativistic proton-nucleus and nucleus-nucleus collisions.  
We observe both an impact parameter dependence and a target mass 
dependence of the strength of the azimuthal correlation function.  
The data are consistent with the geometrical picture of pions 
suffering secondary interactions while traveling through target 
spectator matter and thereby heating up the entire target nucleus.  
The latter argument is supported both by the observed linear rise 
of the target rapidity proton multiplicity with target mass $A$, as 
well as by two-proton correlations [22] where the extracted source 
sizes show a dependence on the target mass similar to $\propto 
A^{1/3}$ and are close to the nuclei radii.

\newpage
\noindent \rule{3cm}{0.01cm}\\ 
$^1$ {\footnotesize Now at University of Karlsruhe, Germany}\\
$^2$ {\footnotesize Institute for Nuclear Studies, Warsaw, Poland}
\normalsize

\frenchspacing
\vspace*{3mm} \noindent
{\bf References:\\}
\small\vspace*{-3ex}
\begin{itemize}
\item[{[1]}]  K.H.~Kampert, J. Phys. G15 (1989) 691 
\item[{[2]}]  H.~Gutbrod, A.M.~Poskanzer, and H.G.~Ritter,
              Rep. Prog. Phys. 52 (1989) 1267  
\item[{[3]}]  H.R. Schmidt, Int. Journal of Mod. Phys. A6 (1991) 3865
\item[{[4]}]  G.F. Chapline, M.H. Johnson, E. Teller, and M.S. Weiss,
              Phys. Rev.D8 (1973) 4302  
\item[{[5]}]  H. St\"{o}cker, W. Greiner, and W. Scheid, Z. Phys.
              A286 (1978) 121  
\item[{[6]}]  H.A.~Gustafsson et al., Plastic Ball Collaboration,
              Phys. Lett. B142 (1984) 141 
\item[{[7]}]  J. Gosset et al., Diog\`{e}ne-Collaboration, Phys.
              Rev. Lett. 62 (1989) 1251  
\item[{[8]}]  H.R. Schmidt et al., WA80-Collaboration,
              Nucl. Phys. A544 (1992) 449c  
\item[{[9]}]  L. Venema et al., TAPS-Collaboration, Phys. Rev. Lett.
              71 (1993) 835  
\item[{[10]}] D. Brill et al., KAOS-Collaboration,
              Phys. Rev. Lett. (1993) 336  
\item[{[11]}] S.A. Bass et al., Phys. Lett. B335 (1994) 289;
              Phys. Rev. C51 (1995) 3343  
\item[{[12]}] J. Barette et al., E877-Collaboration, Nucl. Phys.
              A590 (1995) 259c 
\item[{[13]}] H.H. Gutbrod et al., Plastic Ball Collaboration,
              Phys. Lett. B216 (1989) 267;
              Phys. Rev. C42 (1990) 640  
\item[{[14]}] H.R. Schmidt and H.H. Gutbrod, Proceedings of a NATO
              Advanced Study Institute on The Nuclear Equation of
              State, Pe\~{n}iscola, Spain, NATO ASI Series B: 
              Physics 216B (1989) 51  
\item[{[15]}] A. Baden et al., Plastic Ball Collaboration, 
              Nucl. Inst. and Meth. 203 (1982) 189  
\item[{[16]}] R. Albrecht et al., WA80-Collaboration, Z. Phys. C55 
              (1992) 539 
\item[{[17]}] K.G.R. Doss et al., Plastic Ball Collaboration, Phys. Rev. 
              C37 (1988) 163  
\item[{[18]}] K. Werner, Phys. Rep. 232 (1993) 87 
\item[{[19]}] K.H. Kampert et al., WA80-Collaboration,
              Nucl. Phys. A544 (1992) 183c  
\item[{[20]}] R. Albrecht et al., Phys. Rev. C44 (1991) 2736 
\item[{[21]}] B. Adyasevich et al., Nucl. Phys. B16 (1990) 419c 
\item[{[22]}] T.C. Awes et al., WA80-Collaboration, Z. Phys. C65 
              (1995) 207  
\end{itemize}
\nonfrenchspacing

\end{twocolumn}

\end{document}